\documentclass[12pt,a4paper,final]{iopart}

\usepackage{iopams}
\usepackage{graphicx}
\usepackage[breaklinks=true,colorlinks=true,linkcolor=blue,urlcolor=blue,citecolor=blue]{hyperref}
\usepackage[font=footnotesize,labelfont=bf,
   justification=justified,
   format=plain]{caption}

\begin{document}

\title[]{Unusual Multiple Magnetic Transitions and Anomalous Hall Effect Observed in Antiferromagnetic Weyl Semimetal,  Mn$_{2.94}$Ge (Ge-rich)}

\author{Susanta Ghosh, Achintya Low, Susmita Changdar, Shubham Purwar, and Setti Thirupathaiah$^*$}
\address{Department of Condensed Matter and Materials Physics, S. N. Bose National Centre for Basic Sciences, Kolkata, West Bengal-700106, India}
\ead{$^*$setti@bose.res.in}

\vspace{10pt}
\begin{indented}
\item[]\today
\end{indented}

\begin{abstract}
We report on the magnetic and Hall effect measurements of the magnetic Weyl semimetal, Mn$_{2.94}$Ge (Ge-rich) single crystal. From the magnetic properties study, we identify unusual multiple magnetic transitions below the N$\acute{e}$el temperature of 353 K, such as the spin-reorientation ($T_{SR}$) and ferromagnetic-like transitions. Consistent with the magnetic properties, the Hall effect study shows unusual behavior around the spin-reorientation transition. Specifically, the anomalous Hall conductivity (AHC) increases with increasing temperature, reaching a maximum at $T_{SR}$, which then gradually decreases with increasing temperature. This observation is quite in contrast to the  Mn$_{3+\delta}$Ge (Mn-rich) system, though both compositions share the same hexagonal crystal symmetry.  This study unravels the sensitivity of magnetic and topological properties on the Mn concentration.
\end{abstract}

\section{Introduction}

Topological materials such as the topological insulator (TI), the Dirac semimetal (DSM), and the Weyl semimetal (WSM) are some of the most widely discussed systems in recent days in the class of topological quantum materials. In Dirac semimetals, the nodal points are protected by both time-reversal and inversion symmetry~\cite{PhysRevLett.108.140405,liu2014discovery}. Breaking the time-reversal or inversion symmetry creates a Weyl semimetal~\cite{PhysRevX.5.031013,yang2015weyl}. In this way, we can have two types of Weyl semimetals based on the broken symmetry, such as the non-magnetic Weyl semimetal created by the inversion symmetry breaking~\cite{Ali2014, Huang2015} and the magnetic Weyl semimetal (MWSM) created by breaking the time-reversal symmetry~\cite{yan2017topological}. Interestingly, in the presence of magnetism, the topological materials exhibit exotic quantum phenomena such as the quantum anomalous Hall effect (QAHE), large intrinsic AHE, topological Hall effect (THE), Chern insulating state, etc.~\cite{PhysRevB.101.094404,ning2020recent,deng2020quantum}. Though many magnetic Weyl semimetals have been predicted theoretically, only a few have been experimentally realized, such as Co$_3$Sn$_2$S$_2$ \cite{liu2018giant,morali2019fermi}, Co$_2$MnGa \cite{doi:10.1126/science.aav2327}, Mn$_3$Sn \cite{article}, Fe$_3$Sn$_2$ \cite{PhysRevLett.125.076403} and YbMnBi$_2$ \cite{borisenko2019time}.

Among the MWSMs, noncollinear and coplanar Mn$_3$X (X = Sn and Ge) antiferromagnets have recently received a great deal of attention from the research community due to their distinct magnetic and topological properties despite sharing a similar crystal structure. For instance, Mn$_{3}$Sn shows a magnetic transition within 260–275 K from a high-temperature noncollinear AFM order to a low-temperature spin-spiral structure~\cite{Sung2018, Yan2019, Low2022}. In contrast, no such magnetic transition has been observed to date in Mn$_{3}$Ge down to the lowest possible temperature~\cite{nayak2016large, yamada1988magnetic,PhysRevApplied.5.064009,xu2020finite,chen2021anomalous}. Few other reports demonstrated the topological Hall effect (THE) in addition to the anomalous Hall effect (AHE) in Mn$_{3\delta}$Sn~\cite{Li2019, Low2022}, while Mn$_{3}$Ge is mainly known to show the AHE without THE. Nevertheless,  the anomalous Hall effect found in both systems mainly originated from the nonzero Berry-phase in the momentum space ~\cite{nayak2016large,chen2021anomalous,PhysRevLett.88.207208,nagaosa2010anomalous}. Further, the electronic and magnetic properties of Mn$_{3}$X (X = Sn and Ge) are sensitive to the Mn concentration and crystal growth techniques~\cite{kren1975study,yamada1988magnetic,PhysRevApplied.5.064009}. Mn$_{3}$Ge can have either a hexagonal or tetragonal phase depending on the synthesis procedure and annealing temperature~\cite{ohoyama1961,PhysRevB.83.174103}.

The tetragonal Mn$_{3}$Ge shows a ferrimagnetic ordering at a Curie temperature of T$_C$ $\approx$ 710 K, and the hexagonal Mn$_{3}$Ge shows antiferromagnetic (AFM) ordering at a N$\acute{e}$el temperature of T$_N$ $\approx$ 365-400 K~\cite{nayak2016large,PhysRevApplied.5.064009,yamada1990atomic,PhysRevB.83.174103}. In the hexagonal phase, the neutron diffraction studies revealed an inverse triangular spin-structure with Mn moments aligned within the kagome layers to create a 120° spin-structure of noncollinear antiferromagnetic ordering~\cite{yamada1988magnetic,PhysRevB.97.214402}. Besides the noncollinear AFM ordering, a tiny ferromagnetic moment is observed in these systems due to the geometrical frustration among the Mn magnetic moments within the Kagome lattice network~\cite{yamada1988magnetic,Tomiyoshi1983}. This tiny spontaneous magnetization combined with the nonzero momentum-space Berry phase gives rise to large AHE at low temperatures~\cite{nayak2016large,PhysRevApplied.5.064009}. Generally, these systems are grown with excess Mn, such as Mn$_{3+\delta}$X where $\delta$ varies between 0.2 and 0.4~\cite{Tomiyoshi1982,nayak2016large, yamada1988magnetic,PhysRevApplied.5.064009}. However, recent studies showed that the self-flux technique can grow the Mn$_{3-\delta}$Sn crystals with Sn-rich~\cite{Sung2018, Low2022}.

In this contribution, we report on the successful growth of high-quality  Mn$_{2.94}$Ge (Ge-rich) single crystals, and a thorough study of the magnetic and Hall effect properties. The magnetic property study shows multiple magnetic transitions in this system, such as spin-reorientation ($T_{SR}$) and ferromagnetic-like transitions below the N$\acute{e}$el temperature of 353 K. Also consistent with the magnetic measurements, the Hall effect data shows an unusual behavior around $T_{SR}$. The anomalous Hall conductivity (AHC) increases with increasing temperature up to $T_{SR}$, and then gradually decreases with increasing temperature. This observation is quite in contrast to the Mn$_{3+\delta}$Ge (Mn-rich) system, as in Mn$_{3+\delta}$Ge the AHC gradually decreases with increasing temperature, with the highest AHC obtained at the lowest temperatures~\cite{nayak2016large,PhysRevApplied.5.064009,Wuttke2019}. We argue that the unusual Hall effect properties observed in the studied Ge-rich,  Mn$_{2.94}$Ge, system are driven by the spin-reorientation transition similar to the Mn$_3$Sn. Our study suggests that the magnetic and Hall effect properties of Mn$_{3}$Ge are very sensitive to the Mn concentration present in the system.

\begin{figure}[t]
\centering
	\includegraphics[width=\linewidth]{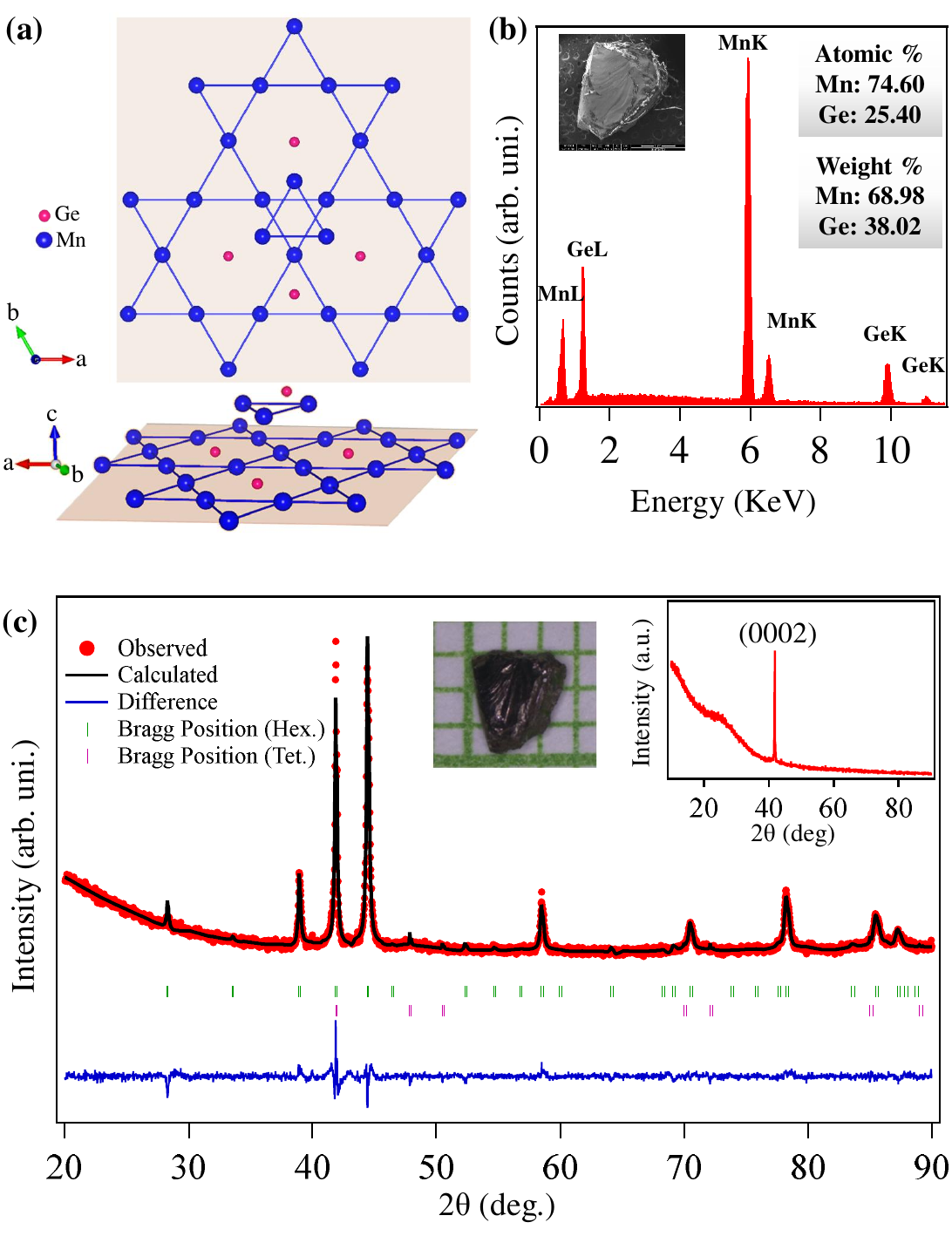}
	\caption{(a) Top view and side view of Mn$_3$Ge hexagonal crystal structure, showing the kagome lattice formed by the Mn atoms. (b) Energy dispersive X-ray spectroscopy (EDS) of Mn$_{2.94}$Ge single crystal. (c) The powder X-ray diffraction pattern of crushed single crystals overlapped with the Rietveld refinement of Mn$_{2.94}$Ge using hexagonal and tetragonal phases. Insets in (c) show a photographic image of Mn$_{2.94}$Ge single crystal and an XRD pattern taken on the single crystal.}
	\label{1}
\end{figure}

\section{Experimental details}
Single crystals of Mn$_{2.94}$Ge were prepared by the melt-growth method using the high-temperature muffle furnace. In this method, Manganese (Alfa Aesar 99.95\%) and Germanium (Alfa Aesar 99.999\%) powders were taken in a ratio of 3.6:1, 
mixed thoroughly in the glove box under an argon environment before sealing in a preheated quartz ampoule in a vacuum of 10$^{–4}$ mbar. The sealed quartz tube was then heated in the muffle furnace up to $1050^{o}$C, and kept at that temperature for the next 24 hours. The tube was slowly cooled to $740^{o}$C at 2 K/h. After prolonged annealing at $740^{o}$C for five more days, the ampoule was quenched in ice water to avoid low-temperature phases. In this method, the obtained Mn$_3$Ge single crystals were of a typical size of 2$\times$2 mm$^2$ and looked shiny.

Phase purity and crystal structure were checked using the powder X-ray diffraction (XRD) technique with  Cu-K$_\alpha$ radiation in a Rigaku X-ray diffractometer (9 KW). Using energy dispersive X-ray spectroscopy (EDS), we find the actual chemical composition of as-prepared samples to be of Mn$_{2.94\pm0.03}$Ge,  which is very close to the nominal composition of Mn$_{3}$Ge. Electrical transport, Hall effect, and magnetic properties studies were performed in a 9-Tesla physical property measurement system (PPMS, Dynacool, Quantum Design) within the temperature range of 2–380 K. Electrical transport and Hall effect measurements were performed using the four-probe technique. Copper leads were attached to the sample using EPO-TEK H21D silver epoxy. For a comparative study, we also grew the single crystals of Mn$_{3.20\pm0.02}$Ge (Mn-rich).

\begin{figure}[ht]
\centering
	\includegraphics[width=0.8\linewidth]{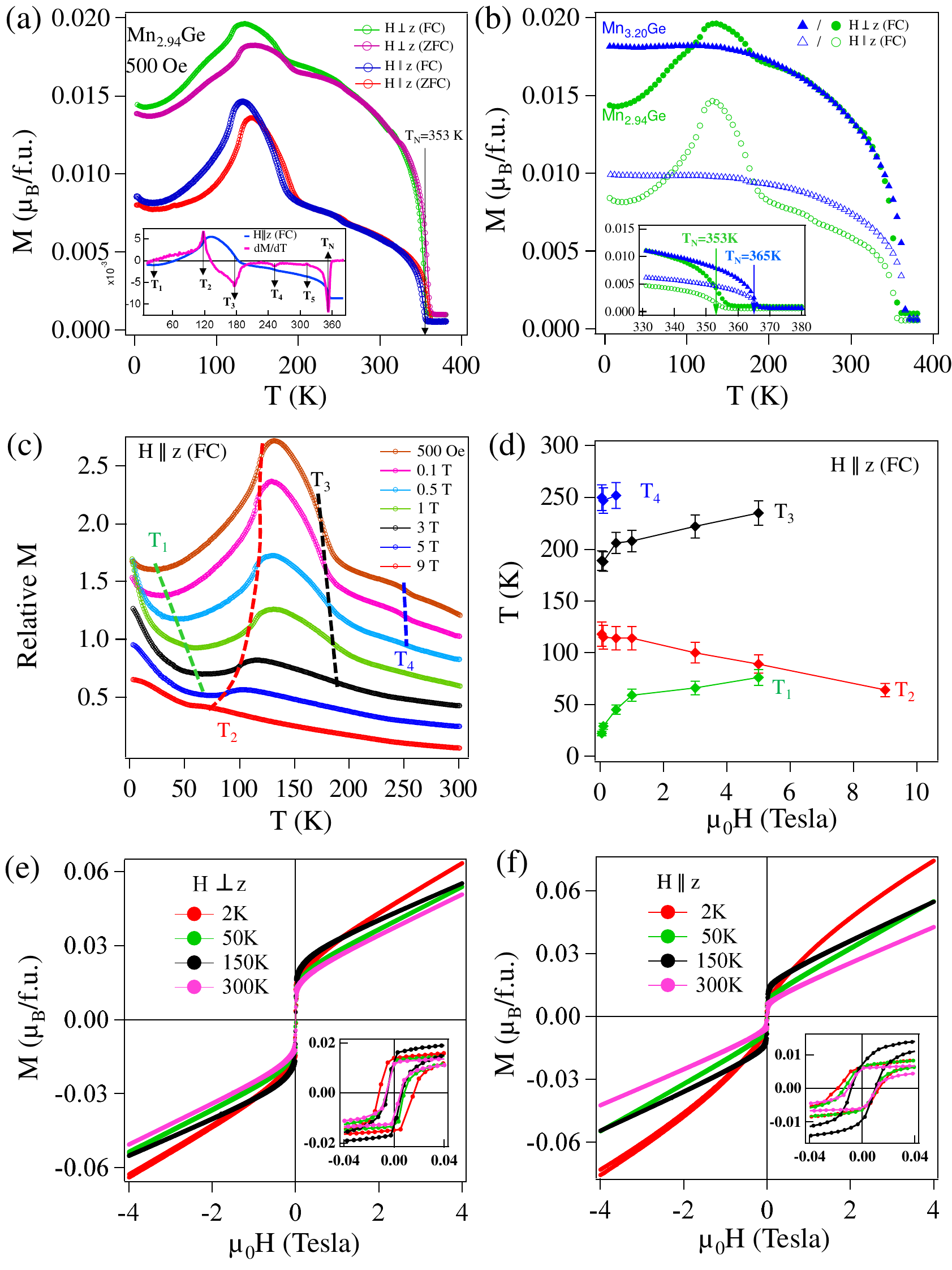}
	\caption{(a) Temperature-dependent magnetization, $M(T)$, measured in zero-field-cooled (ZFC) and field-cooled (FC) modes for $H\parallel z$ and $H\perp z$. Inset in (a) depicts $dM/dT$ of FC $M(T)$ taken for $H\parallel z$, clearly showing multiple magnetic transitions $T_1-T_5$ and a N$\acute{e}$el temperature ($T_N$). (b) Overlapped $M(T)$ data of both Mn$_{2.94}$Ge and Mn$_{3.20}$Ge single crystals. The bottom inset in (b) demonstrates the change in N$\acute{e}$el temperature ($T_N$) with Mn concentration. (c) $M(T)$  measured under various magnetic fields for $H\parallel z$ in FC mode. The dashed lines on the data are the eye-guides showing shifts in the transition temperatures with increasing applied fields, drawn with the help of $dM/dT$ as shown in the supplementary information (see Fig.S1).  (d) Transition temperatures, $T_1$, $T_2$, $T_3$, and $T_4$, are plotted as a function of the applied field. The top-right inset in (d) shows $M(T)$ measured at different fields up to 9 T in the FC mode for $H\parallel z$. (e) and (f) Magnetization isotherms, $M(H)$, measured at different temperatures for $H\perp z$  and  $H\parallel z$, respectively.}
	\label{fig2}
\end{figure}

\section{Results and Discussions}
\subsection{Structural Properties}

Mn$_{3}$Ge is known to crystalize into the Ni$_{3}$Sn-type hexagonal structure with a space group of P6$_{3}$/mmc (194). The Mn atoms form the kagome lattice with triangles and hexagons in the $\it{xy}$-plane, while the Ge atom sits at the centre of the hexagon. A couple of such kagome lattice planes are stacked along the $z$-axis per unit cell, as shown in Fig.~\ref{1}(a). Powder XRD pattern of crushed single crystals is shown in Fig.~\ref{1}(c) overlapped with the Rietveld refinement performed using the hexagonal and tetragonal mixed phases. The Rietveld refinement confirms the majority phase of Mn$_{2.94}$Ge to the hexagonal crystal structure with a small impurity tetragonal phase (3.5\%). Earlier reports too suggested the presence of hexagonal and tetragonal mixed phases in these systems~\cite{PhysRevApplied.5.064009, Rai2022}. Except for the small impurity tetragonal phase, we did not notice any other impurity phases such as Mn$_5$Ge$_3$, Mn$_5$Ge$_2$, Mn$_{11}$Ge$_8$, etc.~\cite{Yamada1986, DucDung2013, Zeng2003} in our studied composition.    Estimated hexagonal lattice parameters from the Rietveld refinement are  $\it{a~(b)}$ = 5.3290(3) $\AA$ and $\it{c}$ = 4.3047(6) $\AA$, which are comparable to the previous reports on Mn$_{3+\delta}$Ge~\cite{xu2020finite,chen2021anomalous}. The XRD pattern taken on the single crystal of Mn$_{2.94}$Ge is shown in the inset of Fig.~\ref{1}(c),  suggesting that the crystal growth plane is parallel to the $\it{z}$-axis.

%\begin{figure*}[ht]
%	\includegraphics[width=1\linewidth]{Fig2.1.pdf}
%\centering
%	\caption{Magnetization isotherms, $M(H)$, measured at different temperatures for (a) $H\perp z$  and (b) $H\parallel z$. The bottom panel and top panel of (c) show spontaneous magnetization ($M_{SP}$) and coercive field ($H_C$) plotted as a function of temperature, respectively.}
%	\label{fig2.1}
%\end{figure*}
\subsection{Magnetic Properties}
Magnetization as a function of temperature [$M(T)$] is measured in field-cooled (FC) and zero-field-cooled (ZFC) modes using a magnetic field ($H$) of 500 Oe applied parallel ($H\parallel z$) and perpendicular ($H\perp z$) to the $\it{z}$-axis as shown in Fig.~\ref{fig2}(a).  The $M(T)$ of Mn$_{2.94}$Ge exhibits an anomalous hump-like structure at around 115 K, clearly visible from both $H\perp z$ and H$\parallel z$ directions. In addition, several unusual magnetic transitions were also identified below the antiferromagnetic transition ($T_N$) of 353 K. To pinpoint the transition temperatures, we performed the first derivative of Magnetization to the temperature ($dM/dT$) as shown in the inset of Fig.~\ref{fig2}(a). Thus, with the help of $dM/dT$, we could identify various unknown transition temperatures $T_1$ to $T_5$ and the AFM N$\acute{e}$el temperature $T_N$. Fig.~\ref{fig2}(b) shows a comparative $M(T)$ data between Mn-deficit Mn$_{2.94}$Ge and Mn-excess Mn$_{3.20}$Ge compounds. Foremost, we observe differing N$\acute{e}$el temperatures between these two compounds such as  T$_N$=365 K for Mn$_{3.20}$Ge and 353 K for Mn$_{2.94}$Ge. Next, we do not observe the low-temperature hump-like structure in the Mn-excess Mn$_{3.20}$Ge, unlike in the Mn-deficit Mn$_{2.94}$Ge.  The N$\acute{e}$el temperature of 365 K in Mn$_{3.20}$Ge is consistent with previous reports on similar Mn-excess Mn$_{3+\delta}$Ge systems. Here, the $T_N$ varies between 365 and 400 K depending on the amount of excess Mn present in the system~\cite{yamada1988magnetic,nayak2016large,PhysRevApplied.5.064009,chen2021anomalous,Wuttke2019}.

To fully explore the nature of unusual magnetic transitions, we measured $M(T)$ in the FC mode by varying the applied field from 500 Oe to 9 T as shown Fig.~\ref{fig2}(c) within the temperature range of 2-300 K.  From the first derivatives (dM/dT) we obtained the transition temperatures $T_1-T_4$ and are plotted in Fig.~\ref{fig2}(d) as a function of applied field. From Fig.~\ref{fig2}(d), we notice that $T_1$ increases with increasing field (from 22 K at 500 Oe to 75 K at 5 T), $T_2$ decreases with increasing field (from 118 K at 500 Oe to 64 K at 9 T), and $T_3$ increases with increasing field (from 188 K at 500 Oe to 235 K at 5 T). These observations clearly suggest that $T_1$ and $T_3$ are ferromagnetic-like transitions, while $T_2$ is an AFM-like transition~\cite{PhysRevLett.106.227206,gupta2014observation,szlawska2020antiferromagnetic}. Further, we find $T_4$ at around 250 K, visible only at lower applied fields ($<$1 T) almost field-independent. Since the field-dependent $M(T)$ was taken only up to 300 K, we are unable to comment on the nature of $T_5$ transition, which is found at 310 K [see Fig.~\ref{fig2}(a)].

Next, the field-dependent magnetization $M(H)$ at different temperatures for the in-plane ($H\perp z$) and out-of-plane ($H\parallel z$) field directions is shown in Figs.~\ref{fig2}(e) and \ref{fig2}(f),  respectively. In-plane and out-of-plane $M(H)$ data display spontaneous magnetization with a small hysteresis loop around the zero field. Linear and unsaturated magnetization at higher fields demonstrates a predominant AFM character, while the spontaneous magnetization and the magnetic hysteresis indicate a weak ferromagnetic component in the system. From the in-plane ($H\perp z$) $M(H)$ data, we estimate the spontaneous magnetization $M_{SP}$ = 0.017$\mu_B$/f.u. and the coercive field $H_C$ =62 Oe at 300 K. From out-of-plane ($H\parallel z$) $M(H)$ data, the values of $M_{SP}$ and $H_C$ are 0.008$\mu_B$/f.u and 111 Oe, respectively, at 300 K. Consistent with the $M(T)$ data, the spontaneous magnetization is higher at 150 K for both field directions [see insets in Figs.~\ref{fig2}(e) and ~\ref{fig2}(f)].

\begin{figure}[ht]
\centering
\includegraphics[width=\linewidth, clip=true]{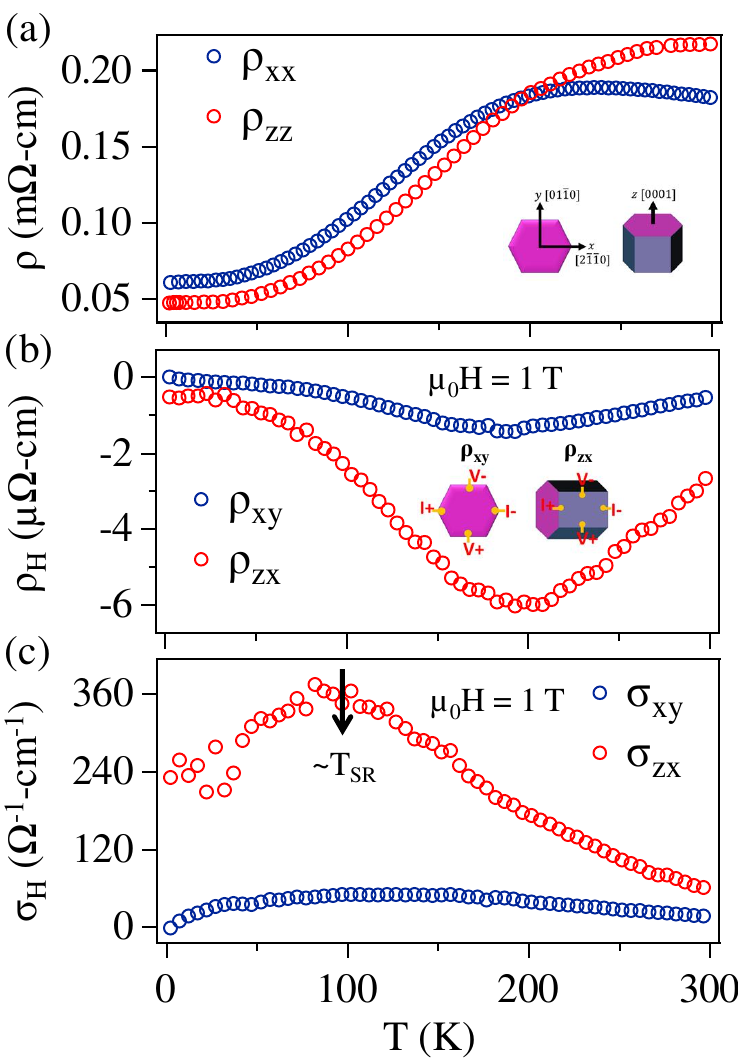}
\caption{(a) Resistivity as a function of temperature plotted for two crystallographic axes directions, $\rho_{xx}$ and $\rho_{zz}$. Inset in (a) shows the $x$,$y$, and $z$-axis directions on the hexagonal crystal. (b) In-plane ($\rho_{xy}$) and out-of-plane ($\rho_{zx}$) Hall resistivity  measured as a function of temperature under a magnetic field of 1 T. Inset in (b) shows Hall resistivity $\rho_{xy}$ and $\rho_{zx}$ measuring geometry.  (c)  In-plane ($\sigma_{xy}$) and out-of-plane ($\sigma_{zx}$) Hall conductivity calculated from the Hall resistivity of (b) plotted as a function of temperature.}
\label{fig3}
\end{figure}

Further, we draw a detailed comparison among our magnetic measurements, Mn-excess Mn$_{3+\delta}$Ge~\cite{Tomiyoshi1982,nayak2016large, yamada1988magnetic,PhysRevApplied.5.064009} and Mn$_{3}$Sn~\cite{Sung2018, Low2022}. Mn$_{3}$Sn usually shows a sudden drop in the magnetization due to spin-reorientation (T$_{SR}$) transition within the temperature range of 265–275 K, depending on the chemical composition and preparation method~\cite{Duan2015, Yan2019}. Note that the spin-reorientation transition in Mn$_3$Sn is from a triangular-spin structure (AFM) to a spiral-spin structure (AFM). Thus, it is an AFM-to-AFM magnetic transition. On the other hand, Mn$_{2.94}$Ge [see Figs.~\ref{fig2}(e) and ~\ref{fig2}(f)] has an AFM ordering throughout the measured temperature range similar to Mn$_3$Sn. Therefore, the gradual decrease of magnetization below $T_2$ in Mn$_{2.94}$Ge is plausibly due to a gradual reorientation of Mn spins [see Fig.~\ref{fig2}(c)], unlike in Mn$_3$Sn where it is spontaneous. Since the low-temperature multiple magnetic transitions below $T_N$=353 K ($T_1$, $T_2$, $T_3$, and $T_4$) are observed only in the Mn-deficit Mn$_{2.94}$Ge but not in the Mn-excess Mn$_{3.20}$Ge, we suggest that the Mn deficiency plays a crucial role in triggering the complex magnetic structure in this system. This suggestion is qualitatively supported by previous reports made on the Mn-deficient Mn$_{3-\delta}$Ge ($\delta$=0.35 and 0.4)~\cite{ohoyama1961, Yamada1982}, where several anomalous magnetic transitions were found below the $T_N$. Further, the spin-reorientation transition temperature $T_{SR}$ observed from different Mn-deficiency systems appears nearly the same. Means, in our studied system of Mn$_{2.94}$Ge the $T_{SR}\approx114$ K, whereas $T_{SR}\approx 120$ K in Mn$_{2.6}$Ge~\cite{Yamada1982}.

\begin{figure}[ht]
\centering
\includegraphics[width=1\linewidth, clip=true]{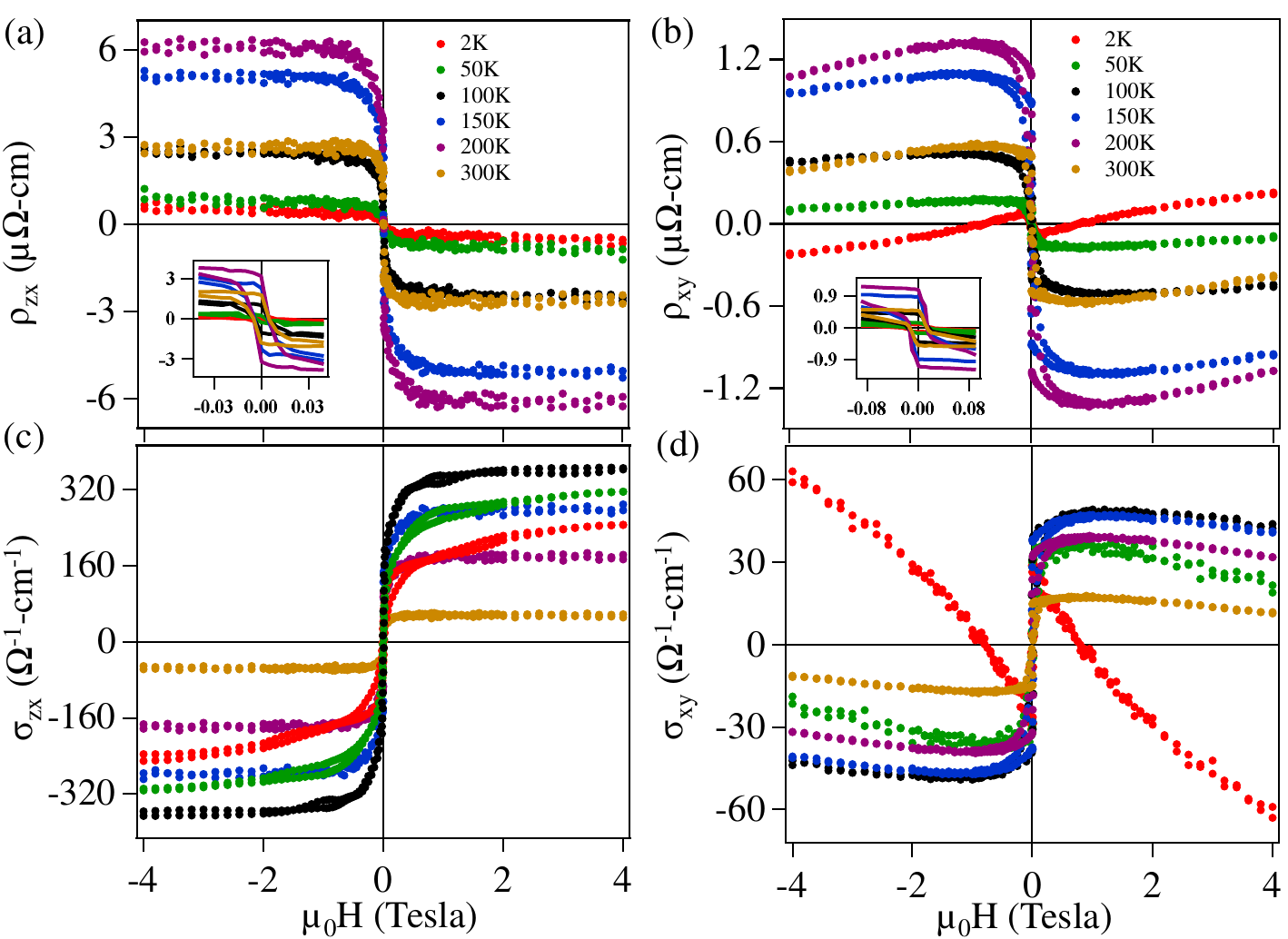}
\caption{(a) and (b) Out-of-plane ($\rho_{zx}$) and in-plane ($\rho_{xy}$) Hall resistivity plotted as a function of applied field, respectively. Zoomed-in data at low magnetic fields are shown in the insets of (a) and (b). (c) and (d) Out-of-plane ($\sigma_{zx}$) and in-plane ($\sigma_{xy}$) Hall conductivity plotted as a function of field calculated from the data of (a) and (b), respectively.}
\label{fig4}
\end{figure}

\subsection{Electrical and Magnetotransport Properties}

In-plane ($\rho_{\it{{xx}}}$) and out-of-plane ($\rho_{\it{{zz}}}$) longitudinal electrical resistivity are plotted as a function of temperature in Fig.~\ref{fig3}(a). Both $\rho_{\it{{xx}}}$ and $\rho_{\it{{zz}}}$ suggest an overall metallic nature of the sample, which is in agreement with previous reports on Mn$_{3+\delta}$Ge~\cite{PhysRevApplied.5.064009,xu2020finite}. Fig.~\ref{fig3}(b) depicts in-plane ($\rho_{\it{xy}}$) and out-of-plane ($\rho_{\it{zx}}$) Hall resistivity plotted as a function of temperature. Here, the Hall resistivity $\rho_{\it{xy}}$ is obtained for the current applied along the $\it{x}$-direction, field applied along the $\it{z}$-direction, and the Hall voltage measured along the $\it{y}$-direction. Similarly, the Hall resistivity $\rho_{\it{zx}}$  is obtained for the current applied along the $\it{z}$-direction, field applied along the $\it{y}$-direction, and the Hall voltage measured along the $\it{x}$-direction. The Hall resistivity was measured using both positive (1 T) and negative (-1 T) magnetic fields and calculated using the formula  $\frac{\rho_H (H)-\rho_H (-H)}{2}$ in order to eliminate the magnetoresistance contribution.  Fig.~\ref{fig3}(c) depicts temperature-dependent in-pane ($\sigma_{xy}$) and out-of-plane ($\sigma_{zx}$) Hall conductivity calculated using the formulae, $\sigma_{xy} = -\frac{\rho_{xy}}{({\rho}^2_{xx} + {\rho}^2_{xy})}$ and $\sigma_{zx} = -\frac{\rho_{zx}}{({\rho}^2_{zz} + {\rho}^2_{zx})}$, respectively.

The values of out-of-plane Hall conductivity ($\sigma_{zx}$) are significantly higher than the values of in-plane Hall conductivity ($\sigma_{xy}$) at all measured temperatures,  consistent with earlier reports~\cite{nayak2016large, PhysRevApplied.5.064009,chen2021anomalous}. As can be noticed from Fig.~\ref{fig3}(c),  $\sigma_{zx}$ increases gradually with decreasing temperature, reaching a maximum of 362.48 $\Omega^{-1}cm^{-1}$ at around 95$\pm$5K. Below this, $\sigma_{zx}$ decreases with decreasing temperature. Similarly, the Hall conductivity $\sigma_{xy}$ increases with decreasing temperature, reaching a maximum of 50 $\Omega^{-1}cm^{-1}$ at around 110$\pm$5K, and below this $\sigma_{xy}$  decreases with decreasing temperature. Notably, the maximum of $\sigma_{xy}$ found at 110 K is close to the spin-reorientation transition ($T_2$ or $T_{SR}$) of 114 K. A similar sort of relation is noticed between the Hall conductivity and spin-reorientation transition of Mn$_3$Sn, where the Hall conductivity gradually increases with temperature down to $T_{SR}$=275 K and suddenly drops to zero below $T_{SR}$~\cite{Sung2018}. Since in the studied system Mn$_{2.94}$Ge, the Mn spins reorient gradually, we should observe a gradual decrease in AHC below the $T_{SR}$. Let us also emphasize here that, although non-zero magnetization ($M$) is necessary to realize the AHE,  the large AHE observed in this system is mainly governed by the electronic band structure originated Berry curvature~\cite{chen2021anomalous, PhysRevLett.88.207208,nagaosa2010anomalous}. Therefore,  it is unlikely that the magnetic transitions at $T_1$, $T_3$, and $T_4$ pose any significant impact on the AHC except for the spin-reorientation transition at $T_{SR}$ as the Berry curvature changes at $T_{SR}$ due to time-reversal symmetry breaking in the presence of triangular spin-structure~\cite{yan2017topological}. Nevertheless, the magnetization [$M(T)$] macroscopically affects the temperature dependence of AHC as evidenced by the similarities in the temperature-dependent $\sigma_{xy}$ [see Fig.~\ref{fig3}(c)] and magnetization [see Fig.~\ref{fig2}(a)]. Precisely, the large AHC intrinsically originated from the Berry curvature is temperature independent, but the temperature dependence of AHC, as observed from Fig.~\ref{fig3}(c), is governed by temperature-dependent magnetization.

%Worth to emphasize here that, in contrast to our observation, previous reports on Mn$_{3+\delta}$Ge showed a continuous increase in Hall conductivity with decreasing temperature before Hall conductivity gets saturated at very low temperatures~\cite{nayak2016large, PhysRevApplied.5.064009,chen2021anomalous}.

\begin{figure}[ht]
\centering
\includegraphics[width=0.8\linewidth, clip=true]{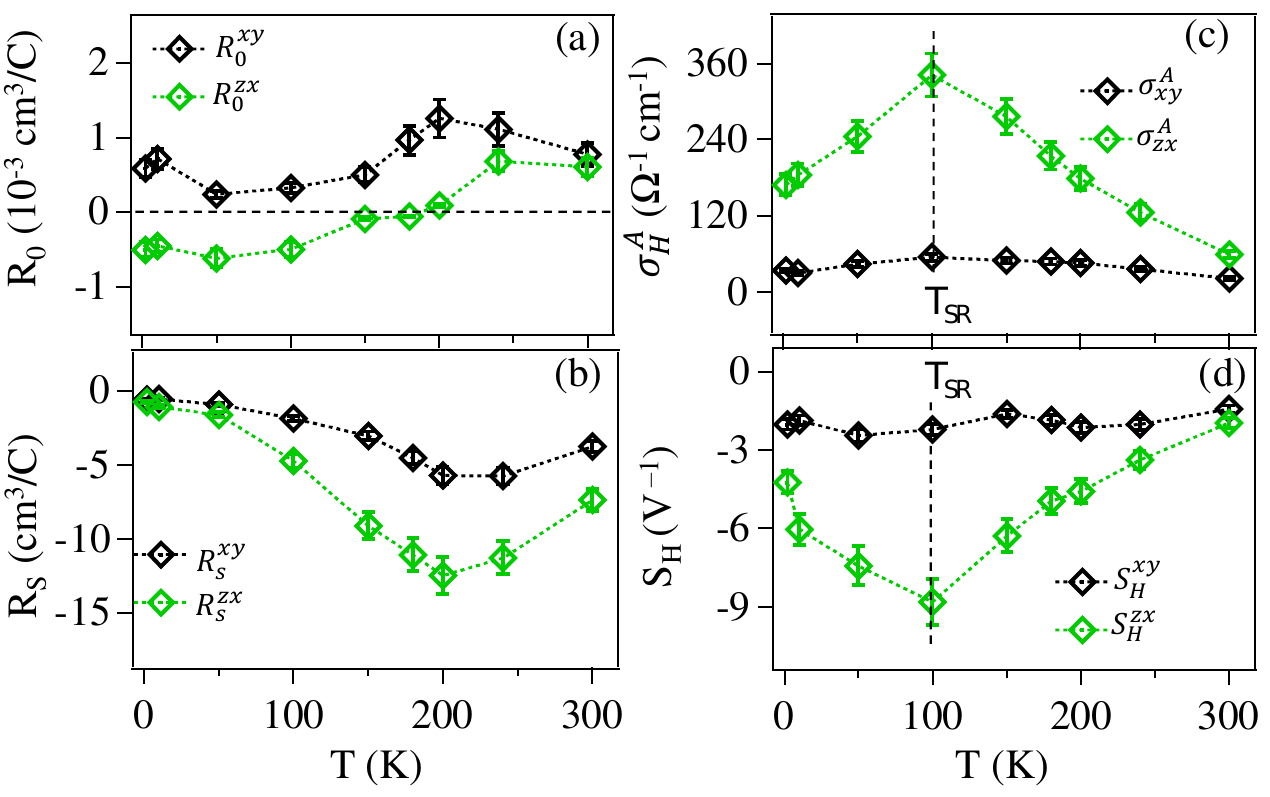}
\caption{Temperature-dependent (a) normal Hall coefficient ($R_0$), (b) anomalous Hall coefficient ($R_s$), (c) anomalous Hall conductivity (${\sigma}^{A}_{H}$), and (d) scaling coefficient ($S_H$).  See the text for more details.}
\label{fig5}
\end{figure}

Next, field-dependent out-of-plane Hall resistivity ($\rho_{\it{{zx}}}$) measured at various temperatures is plotted in Fig.~\ref{fig4}(a). We observe a sudden jump in $\rho_{\it{{zx}}}$ near the zero field, which then saturates with increasing magnetic field, the signature of an anomalous Hall effect. From the zoomed-in data of $\rho_{\it{{zx}}}$ as shown in the inset of Fig.~\ref{fig4}(a), we observe a hysteretic behavior near the zero fields that is consistent with the magnetization hysteresis [see Fig.~\ref{fig2}(e)]. Fig.~\ref{fig4}(c) shows out-of-plane Hall conductivity ($\sigma_{\it{{zx}}}$) plotted as a function of field. In agreement with $\sigma_{\it{{zx}}}(T)$ shown in Fig.~\ref{fig4}(c), we observe a maximum anomalous Hall jump in $\sigma_{\it{zx}(H)}$ at around 100 K. Similarly, field-dependent in-plane Hall resistivity ($\rho_{\it{{xy}}}$) and Hall conductivity ($\sigma_{\it{{xy}}}$) are plotted in Figs.~\ref{fig4}(b) and ~\ref{fig4}(d), respectively. Since the in-plane anomalous Hall jump is small, $\sigma_{\it{{xy}}}$ is mostly dominated by the normal Hall contribution, which varies linearly with the field. Thus, unlike in $\rho_{\it{{zx}}}$, we do not find saturation in $\rho_{\it{{xy}}}$. Further, nonlinear Hall conductivity around zero fields is observed in $\sigma_{\it{{xy}}}$ at 2 K, which is in agreement with the $M(H)$ data [see Fig.~\ref{fig2}(f)]. Since the Hall resistivity shown in Figs.~\ref{fig4}(a) and \ref{fig4}(b) is the total Hall resistivity ($\rho_H$), they have contributions from the normal Hall resistivity (${\rho}^N_H$) and anomalous Hall resistivity (${\rho}^A_H$) in such a way that $\rho_H$ = ${\rho}^N_H$ + ${\rho}^A_H$. Here, ${\rho}^N_H$ = R$_0$$\mu_0$H and ${\rho}^A_H$ = R$_S$$\mu_0M$. R$_0$ is the normal Hall coefficient, and R$_S$ is the anomalous one.

Figs.~\ref{fig5}(a) and \ref{fig5}(b)  depict temperature-dependent R$_0$ and R$_S$ values, respectively,  obtained by fitting $\rho_{\it{{zx}}}$ and $\rho_{\it{{xy}}}$ data using the relation $\rho_H$ = R$_0$$\mu_0$H+R$_S$$\mu_0$M. The value of out-of-plane $R^{zx}_S$ is about three orders higher than $R^{zx}_0$, indicating a clear dominance of the anomalous Hall contribution over the normal Hall contribution in the out-of-plane. Fig.~\ref{fig5}(c) depicts  ${\sigma}^{A}_{H}$  plotted as a function of temperature calculated using the relation $\sigma^{A}_{H}=-\frac{M \mu_0 R_S}{\rho^2}$.  At room temperature, the values of in-plane and out-of-plane anomalous Hall conductivity are of ${\sigma}^A_{zx}$ = 59.09 ${\Omega}^{-1}{cm}^{-1}$ and ${\sigma}^A_{xy}$ = 20.63 ${\Omega}^{-1}{cm}^{-1}$. We observe that the Hall conductivities are maximum, ${\sigma}^A_{zx}$ = 342.05 ${\Omega}^{-1}{cm}^{-1}$ and ${\sigma}^A_{xy}$ = 54.94 ${\Omega}^{-1}{cm}^{-1}$, at around the spin-reorientation transition temperature of 100 K. However, below 100 K, consistent with the magnetization data [see Fig.~\ref{fig2}(a)], ${\sigma}^A_{xy}$ and ${\sigma}^A_{zx}$ decrease with temperature. Further, the anomalous Hall resistivity (AHR) is generally proportional to ${\rho}^2$ as it originates from the intrinsic Berry curvature~\cite{nagaosa2010anomalous,PhysRevApplied.5.064009}. Thus, the scaling coefficient can be written as $S_{H}$ = $\frac{\mu_0R_S}{{\rho}^2}$. Fig.~\ref{fig5}(d) depicts $S_H$ plotted as a function of temperature. We find that $S^{zx}_{H}$ strongly depends on the temperature, $S^{zx}_{H}$$\sim$ -1.95 V$^{-1}$ (at 300K), -8.81 V$^{-1}$ (at 100K), and -4.24 V$^{-1}$ (at 2 K). On the other hand, $S^{xy}_{H}$ is nearly constant at all measured temperatures. In general, for the intrinsic AHE, ${\sigma}^A_{H}$ $\propto$ M, the scaling coefficient $S_H$ should be a temperature-independent constant~\cite{Manyala2004,PhysRevLett.96.037204,PhysRevB.94.075135}. 

\begin{figure}[ht]
\centering
\includegraphics[width=\linewidth, clip=true]{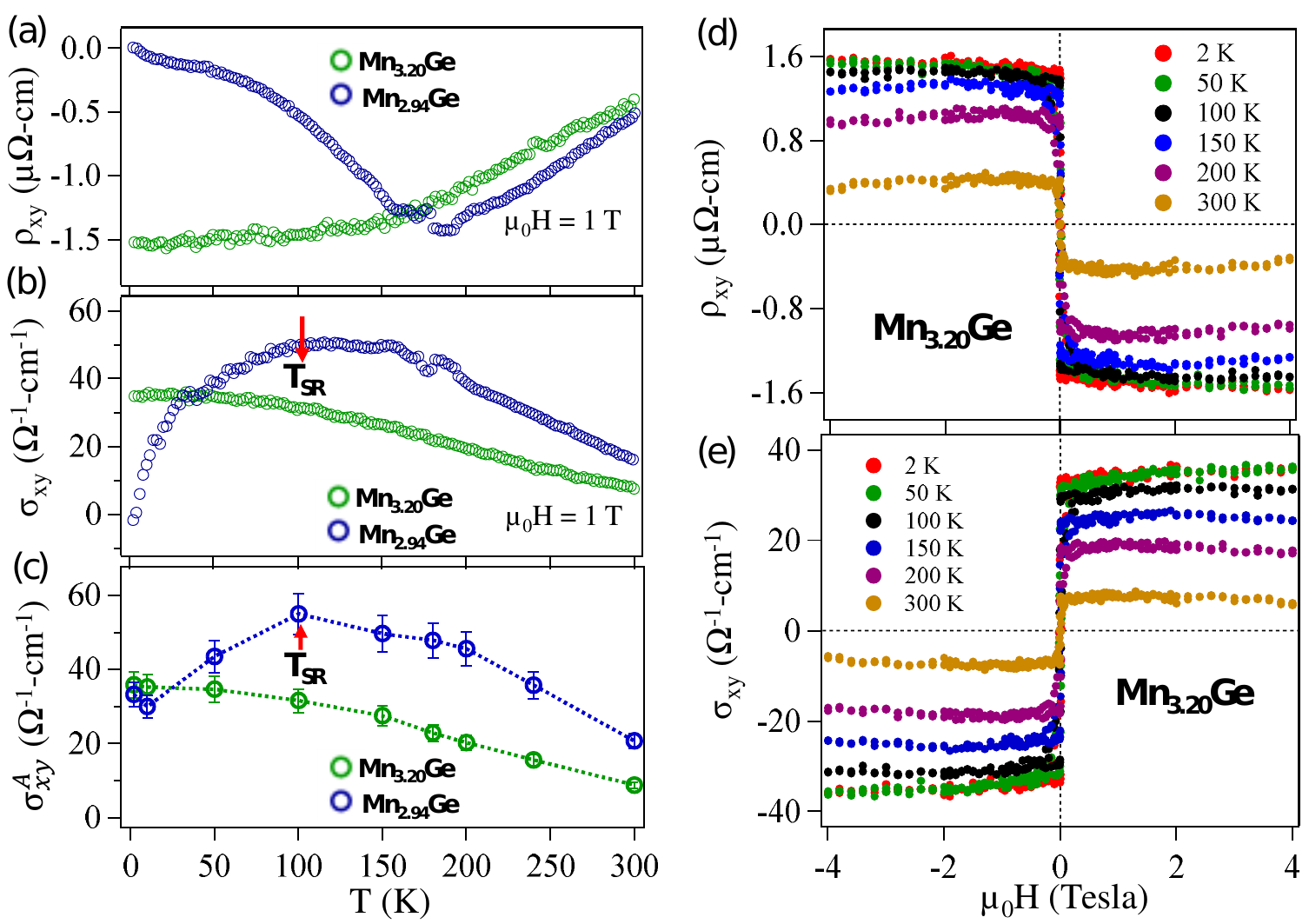}
\caption{(a) In-plane ($\rho_{\it{{xy}}}$) Hall resistivity data as a function of temperature measured with a magnetic field of 1 T from both Mn$_{3.20}$Ge and Mn$_{2.94}$Ge samples. (b) In-plane ($\sigma_{\it{{xy}}}$) Hall conductivity calculated from the Hall resistivity of (a) plotted as a function of temperature. (c) Anomalous Hall conductivity ($\sigma^A_{\it{{xy}}}$), calculated from field dependent Hall resistivity data. (d) and (e) $\rho_{\it{{xy}}}$ and $\sigma_{\it{{xy}}}$ plotted as a function of applied field for Mn$_{3.20}$Ge.}
\label{fig6}
\end{figure}

Finally, in Fig.~\ref{fig6} we compare the Hall effect data between Mn-rich (Mn$_{3.2}$Ge) and Ge-rich (Mn$_{2.94}$Ge) compounds. Figs.~\ref{fig6}(a) and ~\ref{fig6}(b) depict the in-plane Hall resistivity ($\rho_{\it{{xy}}}$) and Hall conductivity ($\sigma_{\it{{xy}}}$) of Mn$_{3.2}$Ge plotted as a function of temperature,   measured with an applied field of 1 T, and overlapped with the similar data of Mn$_{2.94}$Ge taken from Figs.~\ref{fig3}(b) and ~\ref{fig3}(c). $\rho_{\it{{xy}}}$ and $\sigma_{\it{{xy}}}$ of Mn$_{3.2}$Ge are in very good agreement with previous reports on Mn-rich, Mn$_{3+\delta}$Ge, systems~\cite{nayak2016large,PhysRevApplied.5.064009,Wuttke2019}. From a closer observation of the Hall conductivity data [see Fig.~\ref{fig6}(b)], we can clearly see that the Hall conductivity of Mn$_{3.2}$Ge monotonically increases with decreasing temperature and reaches maximum $\sigma_{\it{{xy}}}$ as we approach the low temperatures. On the other hand, in the case of Mn$_{2.94}$Ge, the Hall conductivity gradually increases with decreasing temperature but a change in the slope is noticed at around $T_{SR}$. Further decreasing the sample temperature below $T_{SR}$, in contrast to Mn$_{3.2}$Ge,  Hall conductivity decreases with temperature. Similar behaviour is noticed from the anomalous Hall conductivity ($\sigma^A_{\it{{xy}}}$) plotted as a function of temperature as shown in Fig.~\ref{fig6}(c).  Figs.~\ref{fig6}(d) and ~\ref{fig6}(e) display $\rho_{\it{{xy}}}$ and $\sigma_{\it{{xy}}}$ of Mn$_{3.2}$Ge plotted as a function of applied field measured at various sample temperatures.  Consistent with the $\rho_{\it{{xy}}}(T)$ and $\sigma_{\it{{xy}}}(T)$ shown in Figs.~\ref{fig6}(a) and ~\ref{fig6}(b), the saturated value of Hall conductivity at higher fields gradually increases with decreasing the temperature. Whereas from the similar data of Mn$_{2.94}$Ge, shown in  Figs.~\ref{fig4}(b) and ~\ref{fig4}(d), we observe gradual decrease in the Hall conductivity above and below $T_{SR}$ with a maximum noticed at $T_{SR}$.

\section{Summary}

In summary,  we have grown Mn$_{2.94}$Ge (Ge-rich) single crystals to study the electrical transport, magnetic, and magnetotransport properties. Importantly, we show that the magnetic and magnetotransport properties of  Mn$_{2.94}$Ge are different from Mn$_{3+\delta}$Ge (Mn-rich), particularly at low temperatures. We identify that the magnetic and Hall properties of Mn$_{2.94}$Ge are qualitatively similar to those of Mn$_3$Sn. Consistent with the magnetic properties, the Hall effect study shows unusual behavior around the spin-reorientation transition. This observation contrasts the previous studies on Mn$_{3+\delta}$Ge as no such spin-reorientation transition is observed. Further, by comparing the results of Mn$_{2.94}$Ge and Mn$_{3.20}$Ge from this study with those of previous reports on Mn$_{3+\delta}$Ge we propose that the Mn concentration plays a crucial role in shaping the magnetic and Hall properties of Mn$_3$Ge.

\section{Acknowledgement}

S.G. acknowledges University Grants Commission (UGC), India for the Ph.D. fellowship. Authors thank SERB (DST), India for the financial support (Grant no. SRG/2020/00393). Part of this work has been done using instruments from the Technical Research Centre (TRC)
of S. N. Bose National Centre for Basic Sciences, established under the TRC project of Department of Science and Technology (DST), Govt. of India.

\section*{References}
\bibliographystyle{iopart-num}
\bibliography{Mn3Ge}% Produces the bibliography via BibTeX.

\end{document}